\documentstyle[pre,aps,epsf,floats]{revtex} 

\begin{document}

\def\bottomfraction{0.5}

\textheight24.6cm

\flushbottom

\twocolumn[\hsize\textwidth\columnwidth\hsize\csname @twocolumnfalse\endcsname

\title{Moment analysis of the probability distributions
of different sandpile models}

\author{S. L\"ubeck\cite{SvenEmail} }

\address{
Theoretische Tieftemperaturphysik, 
Gerhard-Mercator-Universit\"at Duisburg,\\ 
Lotharstr. 1, 47048 Duisburg, Germany \\}

\date{Received 1 April 1999; revised manuscript received 26 August 1999}

\maketitle

\begin{abstract}
We reconsider the moment analysis of the Bak-Tang-Wiesenfeld
and the Manna sandpile model in two and three dimensions.
In contrast to recently performed investigations our 
analysis turns out that the models are characterized
by different scaling behavior, i.e., they belong
to different universality classes. 
\end{abstract}


]  

\setcounter{page}{1}
\markright{\rm accepted for publication in Physical Review E.}
\thispagestyle{myheadings}
\pagestyle{myheadings}

\section{Introduction}

The Bak-Tang-Wiesenfeld (BTW) model was introduced as
a paradigm of the concept of self-organized
criticality which describes the emergence of spatiotemporal
correlations in slowly driven dissipative 
systems~\cite{BAK_1,BAK_2}.
Despite its analytical tractability~\cite{DHAR_2}
the scaling behavior of the two-dimensional
BTW model is not well understood.
Especially the exponents which determines
the avalanche distributions are not 
known exactly.
Several numerical attempts were made but
do not provide consisting  
results~\cite{MANNA_1,GRASS_1,CHRIS_2,ROBINSON_2,BENHUR_1,LUEB_2,EDNEY_1}.
Recently \mbox{De\,Menech} {\it et~al.}~performed
a moment analysis of the BTW model~\cite{DEMENECH_1}
which was extended by several authors
to different sandpile 
models~\cite{CHESSA_2,VAZQUEZ_1,CHESSA_3}.
Especially the moment analysis of the size 
distribution of the BTW and  Manna 
sandpile model has led Chessa {\it et~al.}~to
the conclusion that both models 
are characterized by the same scaling exponents
and thus belong to the same universality class~\cite{CHESSA_2}.
In this work we reconsider the moment
analysis and compare the scaling behavior
of various avalanche quantities for
the BTW and Manna model.
Our analysis turns out that in contrast 
to~\cite{CHESSA_2} the moment behavior
of both models differs significantly, 
i.e., the BTW and the Manna model belong
to different universality classes.

\section{Models and simulations}

The BTW model is defined on a $D$-dimensional square lattice
of linear size~$L$ in which non-negative integer variables~$E_{\bf r}$
represent a physical quantity such as the local energy,
stress, height of a sand column, etc.
One perturbes the system by adding particles at a randomly
chosen site $\bf r$ according to
\begin{equation}
E_{\bf r} \, \mapsto \, E_{\bf r}+1\, , \hspace{0.6cm} 
\mbox{with random }{\bf r}.
\label{eq:perturbation}
\end{equation}
A site is called unstable if the corresponding variable
exceeds a critical value $E_{\rm c}$, i.e., if 
$E_{\bf r} \ge E_{\rm c}$, where the critical value is given
by $E_{\rm c}=2 D$.
An unstable site relaxes, its value is decreased by
$E_{\rm c}$ and the $2 D$ nearest neighboring sites are
increased by one unit, i.e.,
\begin{equation}
E_{\bf r}\;\to\;E_{\bf r}\,-\,E_{\rm c}
\label{eq:relaxation_1}
\end{equation}
\begin{equation}
E_{nn,\bf r}\;\to\;E_{nn,\bf r}\;+\;1.
\label{eq:relaxation_2}
\end{equation}

In this way the neighboring sites may be activated and 
an avalanche of relaxation events may take place.
These avalanches are characterized by several physical
properties like the size $s$~(number of relaxation events), 
the area $a$~(number of distinct toppled sites), the
time $t$~(number of parallel updates until the configuration
is stable), the radius $r$ (radius of gyration), the
perimeter~$p$ (number of boundary sites), etc.
In the critical steady state the corresponding probability
distributions should obey power-law behavior~\cite{BAK_1}
\begin{equation}
P_x (x) \; \sim \; x^{-\tau_x}
\label{eq:power_law}
\end{equation}
characterized by the avalanche exponents $\tau_x$ with 
$x\in\{s,a,t,r,p\}$.
Assuming that the size, area, etc.~scale as 
power of each other, 
\begin{equation}
x \sim {x'}^{\gamma_{xx'}}
\label{eq:scaling_ansatz}
\end{equation}
one obtains the scaling 
relations~$\gamma_{xx'}=(\tau_{x'}-1)/(\tau_x -1)$.
The scaling exponents $\gamma_{xx'}$ describe
the static avalanche properties
as well as its propagation.
For instance, the exponent $\gamma_{sa}$ indicates
if multiple toppling events are relevant ($\gamma_{sa}>1$)
or irrelevant ($\gamma_{sa}=1$). 
The exponent $\gamma_{ar}$ equals the fractal
dimension of the avalanches.
A possible fractal behavior of the avalanche boundary 
corresponds to the inequality $D-1<\gamma_{pr}<D$.
Finally, the exponent $\gamma_{tr}$ is usually identified
with the dynamical exponent~$z$.

\begin{figure}[t]
 \epsfxsize=8.0cm
 \epsfysize=6.8cm
 \epsffile{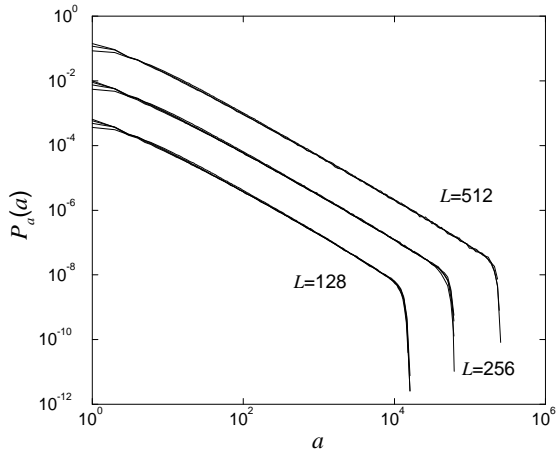}
 \caption{The probability distribution $P_a(a)$ of the
	  two-dimensional Manna model for
	  $E_{\rm c}\in\{2,3,5,10\}$ and various system sizes~$L$.
         For $L<512$ the curves are shifted in the downward 
	  direction. 
	  No significant difference between the curves for
         different values of the critical energy can be observed,
         i.e., $E_{\rm c}$ does not affect the scaling behavior of
	  model.
 \label{fig:manna_2d_hc}} 
\end{figure}

A stochastic version of the BTW model was introduced
by Manna~\cite{MANNA_2}.
Here, critical sites relax to zero, i.e.,
$E_{\bf r}\;\to\;0$
if $E_{\bf r}\ge E_{\rm c}$
and the removed energy is randomly distributed
to the nearest neighbors in the way that one 
chooses randomly for each energy unit one neighbor.
For $E_{\rm c}=1$ the behavior of the model
corresponds to a simple random walk.
Above this value~($E_{\rm c}\ge 2$) is 
the choice of the critical energy irrelevant
to the scaling 
behavior (see Fig.~\ref{fig:manna_2d_hc}).

Recently Dhar introduced a modified version of the 
two-dimensional Manna model where the energy of critical sites
is not reduced to zero but $E_{i,j} \to E_{i,j} - 2$. 
The energy $\Delta E =2$ is then equally distributed with 
probability $1/2$ to the sites ($i\pm1,j$) or otherwise to the 
sites ($i,j\pm1$)~\cite{DHAR_7}.
In this case it is possible to extend an operator
algebra, which was successfully applied in studying the 
BTW model~\cite{DHAR_2}, to this modified Manna model.

Compared to the BTW model the dynamics of the Manna model with its
stochastic distribution of the energy to the nearest neighbors
can be interpreted as a disorder effect.
A different kind of disorder effects were investigated
in directed sandpile models by introducing stochastic 
toppling conditions.
In particular the exact solution of the directed 
BTW model~\cite{DHAR_1} was used as a starting-point 
in order to examine how the known scaling behavior of 
the system is affected by the stochastic toppling rules.
Depending on the details of the introduced disorder a collapse 
of the critical behavior~\cite{TADIC_3}, 
non-universal critical behavior~\cite{LUEB_1,TADIC_1}
as well as a crossover to the different universality class
of directed percolation~\cite{TADIC_2} was observed.

Thus an important question is if the scaling behavior 
of the Manna model differs from the scaling behavior of the 
BTW model, i.e., if the additional fluctuations of
the Manna model could change the universality class.
A real-space renormalization scheme predicted that both 
models belong to the same universality
class~\cite{PIETRO_3,VESPIGNANI_3}.
Here, the authors used a mean-field-type 
approximation~\cite{KATORI_1} in order to perform a 
block transformation.
Therefore it is not clear if this renormalization
ansatz is an appropriate tool to take
the additional fluctuations of the energy 
distribution of the Manna model into account. 

A momentum space analysis of Langevin equations 
of the BTW and the related
Zhang model~\cite{ZHANG_1} predicted that both models
are characterized by the same 
exponents~\cite{DIAZ_1,CORRAL_1}
and numerical investigations confirmed this 
prediction~\cite{BENHUR_1,LUEB_3}.
Unfortunately up to now no extension of the
momentum space analysis to the Manna model could be performed.
The crucial point of this renormalization 
approach is the choice of an appropriate
noise-correlator~\cite{CORRAL_1}.
Compared to the BTW model the corresponding 
analysis of the Manna model requires a different 
noise-correlator.
But a different noise-correlators could lead to
a different critical behavior~(see for instance~\cite{DIAZ_3}).

The first numerical indication that both models
belong to different universality classes was
reported by Ben-Hur and Biham~\cite{BENHUR_1}.
They measured several scaling exponents~$\gamma_{xx'}$
via Eq.~(\ref{eq:scaling_ansatz}) and found,
in particular,
that the dynamical exponents~$z$, the multi-toppling
exponents~$\gamma_{sa}$ and the boundary 
exponents~$\gamma_{pr}$ of both models differ significantly.
But one has to mention that at least one of
the scaling exponents of the BTW model~($\gamma_{sa}$) 
cannot be determined in this way since the assumed
scaling behavior~$s\sim a^{\gamma_{sa}}$ is not 
well defined~\cite{LUEB_2,CHESSA_2}.

However, the conjecture of Ben-Hur and Biham was confirmed
for $D=2$ by a numerical determination of the avalanche
exponents~$\tau_s$, $\tau_a$, $\tau_t$, and $\tau_r$~\cite{LUEB_2}
which again differ significantly for both models.
In this analysis the exponents of the Manna
model were obtained by a direct regression analysis as well as
a simple finite-size scaling analysis
of the corresponding probability distributions.
In the case of the BTW model it was found that
the probability distributions are affected by
unconventional logarithmic finite-size 
corrections~\cite{MANNA_1,LUEB_2} which lead to uncertain 
results for the simple finite-size
scaling ansatz.
Taking these corrections into account it is
possible to estimate the values of the 
exponents~\cite{MANNA_1,LUEB_2} by an extrapolation
to $L\to \infty$.
But one has to note that the assumed logarithmic corrections
are found only numerically, i.e., there exist up to now 
no analytical justification of this unconventional behavior.

These difficulties vanish in the three-dimensional case
where both models fulfill the finite-size scaling 
ansatz
which makes the analysis much easier~\cite{LUEB_4}.
The accuracy of the determination is sufficient to show
that both models belong to different universality classes~\cite{LUEB_4}.
Additionally the scaling behavior of the 
three-dimensional Manna model is strongly affected by
multiple toppling events ($\tau_s\ne \tau_a$) whereas 
it seems that the rare multiple toppling events of 
the BTW model does not contribute to the scaling 
behavior~\cite{GRASS_1,LUEB_4}.
Taking this results into consideration we have convincing,
but of course not completely rigorous, arguments that the BTW and 
the Manna model does not belong to the same universality class.

Recently this statement was questioned by 
Chessa~{\it et al.}~\cite{CHESSA_2}
who performed a moment analysis of both models in
analogy to the investigations
of De\,Menech~{\it et~al.}~\cite{DEMENECH_1}.
The $q$-moment of the probability distribution~$P_x(x)$
is defined as 
\begin{equation}
\langle x^q \rangle \; = \; \int \, dx\, x^q \, P_x(x).
\label{eq:def_q_moment}
\end{equation}
The finite system size $L$ causes a cut-off of the probability
distribution at $x_{max} \sim L^{\gamma_{xr}}$.
Assuming a power-law behavior of the 
distributions~[Eq.\,(\ref{eq:power_law})] the scaling behavior
of the $q$-moment is dominated by the upper
boundary of Eq.~(\ref{eq:def_q_moment})
\begin{equation}
\langle x^q \rangle_L \; \sim \; L^{\sigma_x(q)}
\label{eq:q_moment_L}
\end{equation}
if $q>\tau_x-1$ and where the moment 
exponent~$\sigma_x(q)$ is given by
\begin{equation}
\sigma_x(q)={\gamma_{xr}(q+1-\tau_x)}.
\label{eq:sigma_q}
\end{equation}
For $q<\tau_x-1$ the moment exponent~$\sigma_x(q)$ behaves
nonlinear with respect to $q$. 
The normalization of the probability distributions 
results in $\sigma_x(q)=0$ for $q\to 0$.
Performing numerical investigations one can obtain
the behavior of the exponent~$\sigma_x(q)$ via 
a regression analysis of $\ln{ \langle x^q \rangle_L}$
as a function of $\ln{L}$, or via 
the logarithmic derivative
\begin{equation}
\sigma_x(q) \; = \; \frac{\partial \ln{\langle x^q \rangle_L}}{\partial \ln{L}}.
\label{eq:log_derivative}
\end{equation}

\begin{figure}[t]
 \epsfxsize=8.0cm
 \epsffile{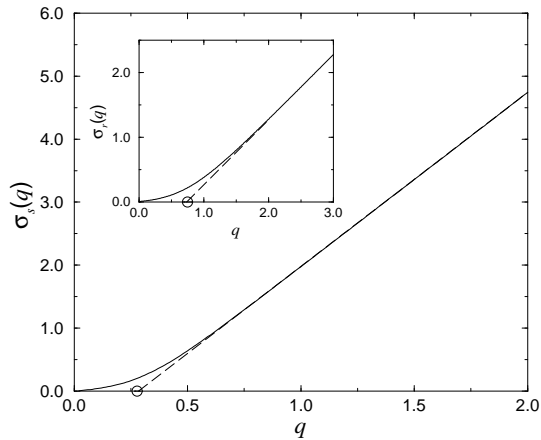}
 \caption{The exponent $\sigma_s(q)$ and $\sigma_r(q)$ 
	  of the two-dimensional Manna model.
	  The extrapolation to the horizontal axis yields
	  the exponent $\tau_s=1.286$ and $\tau_r=1.729$, 
	  respectively.
	  The slope corresponds to the scaling exponents 
	  $\gamma_{sr}$ and $\gamma_{rr}$ where the latter
	  equals one (see Table~\protect\ref{table:gamma}).
	  The positions and sizes of the circles on the horizontal 
	  axes correspond to the values and error-bars
          of the exponents~$\tau_s=1.275\pm 0.011$ and $\tau_r=1.743\pm 0.025$, 
	  obtained from a regression analysis~\protect\cite{LUEB_2}.
 \label{fig:manna_sigma_2d}} 
\end{figure}

The behavior of $\sigma_x(q)$ as a function of $q$ allows
to determine the scaling exponent~$\gamma_{xr}$, which corresponds
to the slope, and the avalanche exponent~$\tau_x$ which 
can be obtained from an extrapolation to the horizontal
axis [Eq.~(\ref{eq:sigma_q})].
This is shown in Fig.~\ref{fig:manna_sigma_2d} 
where the exponents~$\sigma_s(q)$ and $\sigma_r(q)$ of 
the Manna model are plotted as a function of $q$.
The values of the exponents are in agreement with those 
of previous investigations obtained from a regression analysis
and a finite-size scaling analysis, respectively~\cite{LUEB_2}.
But one has to mention that the accuracy of the regression
analysis is higher than the accuracy of the moment analysis.
In the case of the regression analysis one obtains the
exponents~$\tau_x$ by a direct fit to the distribution.
Whereas for the moment analysis one has first to calculate
the average~[Eq.\,(\ref{eq:def_q_moment})],
second to fit the logarithmic derivatives 
[Eq.\,(\ref{eq:log_derivative})] 
and third to extrapolate to
$\sigma_x(q)=0$. 
This leads to a propagation of errors
which increases the uncertainty significantly.

The $q$-dependence of the moment exponent~$\sigma_x(q)$
is determined by the avalanche exponents [Eq.~(\ref{eq:sigma_q})].
Therefore, the moment analysis can be used 
to distinguish the universality classes of different models.
Using the scaling relation 
$\gamma_{xr}=(\tau_r-1)/(\tau_x-1)$, 
Eq.~(\ref{eq:sigma_q}) now reads
\begin{equation}
\sigma_x(q)=\gamma_{xr} \, q \, + \, \Sigma
\label{eq:sigma_q_Sigma}
\end{equation}
with $\Sigma = 1- \tau_r$.
Thus, we get that the intercept~$\Sigma$ of 
the linear $q$-dependence of the moment 
exponent~$\sigma_x(q)$ is the same for
all distributions (site, area, duration, etc.) and
is therefore a characteristic quantity of the 
model.
Considering two models we get that different
values of $\Sigma$ implies different universality
classes.
But the same value of the intercept $\Sigma$
does not imply that both models belong to the
same universality class. 
It is possible that the models display different values
of $\gamma_{xr}$ which results in different
values of $\tau_x=1-\Sigma/\gamma_{xr}$.
Thus two different models belong to the same
universality class if they are characterized by the 
same linear dependence of $\sigma_x(q)$ for all
relevant quantities $x$.

The determination of the intercept~$\Sigma$ 
for various distributions allows to estimate the
accuracy of the moment analysis.
In the case of the Manna model we obtain from the moment analysis
of the size, area, duration and radius distribution
the values
$\Sigma_s=-0.7900 \pm 0.002$,
$\Sigma_a=-0.7202 \pm 0.003$,
$\Sigma_t=-0.7684 \pm 0.004$,
$\Sigma_r=-0.7333 \pm 0.005$.
The corresponding values of the scaling exponents
$\gamma_{xr}$ are listed in Table~\ref{table:gamma}.
The above error-bars correspond to the uncertainty
of the linear regression~[Eq.~(\ref{eq:sigma_q_Sigma})].
The average value 
$\Sigma_{{\rm Manna},2D}=0.7530 \pm 0.037$
agrees with $\Sigma=3/4$ predicted in~\cite{LUEB_2}.
The latter error-bar reflects the uncertainty of the
whole method and could be use as a lower bound for
the error of the avalanche exponent $\Delta \tau_r \ge 0.037$.
Typical error-bars of a direct analysis of
the probability distributions are $\Delta \tau_r \lesssim 0.025$.

\section{Moment analysis of the BTW and Manna model}

\begin{figure}[b]
 \epsfxsize=8.0cm
 \epsffile{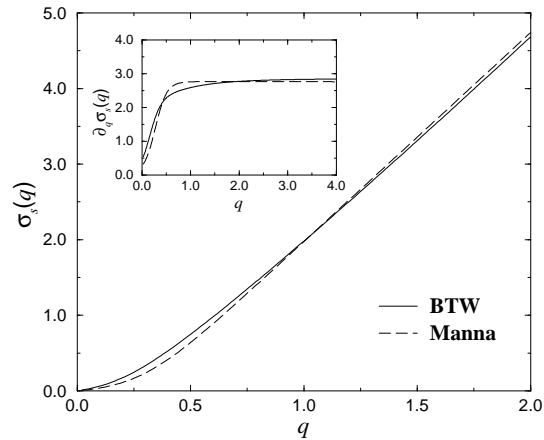}
 \caption{The exponent $\sigma_s(q)$ of the two-dimensional 
          BTW and Manna model.
	  The curves differs slightly.
	  The derivative of both curves is plotted in the
	  inset and reveals that the exponent $\sigma_s(q)$
	  of the BTW model is not well defined in the sense
          that no clear saturation of the exponent can be
	  seen for $q>1$.
 \label{fig:btw_manna_sigma_s_2d}} 
\end{figure}

We now consider the moment analysis of 
the BTW and Manna model for the size, area, duration, radius, and 
perimeter distribution and compare the 
corresponding results.
In the case of the BTW model we analyze  
the probability distributions 
of lattice sizes up
to $L=4096$ for $D=2$ and $L=256$ for $D=3$, respectively.
Since the Manna model does not display strong 
finite-size effects as the BTW model 
it is sufficient to consider system sizes only
up to $L=2048$ for $D=2$.

\begin{figure}[b]
 \epsfxsize=8.0cm
 \epsffile{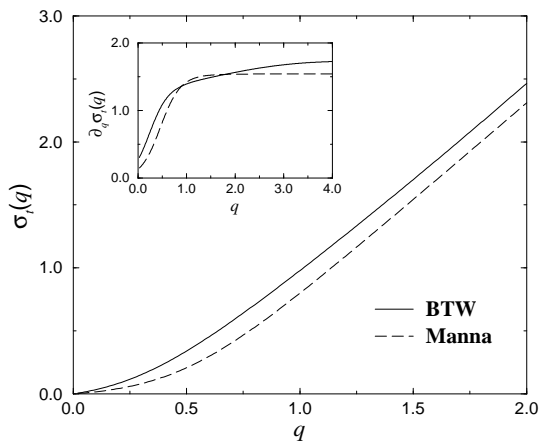}
 \caption{The exponent $\sigma_t(q)$ of the two-dimensional 
          BTW and Manna model.
	  The derivative of both curves is plotted in the
	  inset and reveals that the exponent $\sigma_t(q)$
	  of the BTW model 
	  displays the same complicated behavior as the 
	  exponent $\sigma_s(q)$.
 \label{fig:btw_manna_sigma_t_2d}} 
\end{figure}

The moment analysis of the size distribution leaded
Chessa~{\it et~al.}~to the conclusion that both
models are characterized by the same scaling behavior~\cite{CHESSA_2}.
Therefore we first consider the size distribution 
and the corresponding exponent $\sigma_s(q)$ is
shown in Fig.~\ref{fig:btw_manna_sigma_s_2d}.
In contrast to Chessa~{\it et~al.} who found that
both models display indistinguishable curves
for $q> 1$ we get slightly different curves.
Plotting the derivative of the exponents~$\sigma_s(q)$
with respect to~$q$ the differences become
more significant~(see inset of Fig.~\ref{fig:btw_manna_sigma_s_2d}).
In the case of the Manna model the derivative of the exponent
saturates quickly with increasing~$q$.
Whereas the derivative~$\partial_q \sigma_s(q)$
for the BTW model is characterized by a finite
curvature, i.e.,~the exponent~$\gamma_{sr}$ is
not well defined in the case of the BTW model. 
The duration probability distribution displays
a similar behavior as it can be seen in 
Fig.~\ref{fig:btw_manna_sigma_t_2d}.
Again the exponent $\sigma_t(q)$ of the BTW model is characterized
by a finite curvature. 

\begin{figure}[t]
 \epsfxsize=8.0cm
 \epsffile{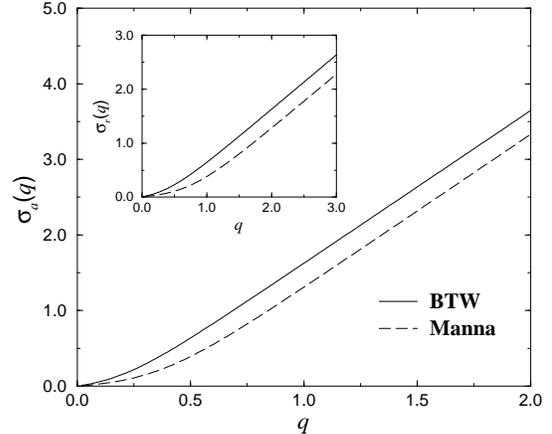}
 \caption{The exponent $\sigma_a(q)$ and $\sigma_r(q)$ 
	  (inset) of the two-dimensional BTW and Manna model.
	  One obtains for both models the same asymptotic
	  slopes (see Table~\protect\ref{table:gamma})
	  which agree with $\gamma_{ar}=2$ and $\gamma_{rr}=1$.
	  But the extrapolation to the horizontal
	  axis yields different values for the models
         indicating that the BTW and the Manna model are
	  characterized by different avalanche exponents
	  $\tau_a$ and $\tau_r$, respectively.
 \label{fig:btw_manna_sigma_a_r_2d}} 
\end{figure}

The $\sigma$-exponents of the radius and area distribution are
plotted for both models in Fig.~\ref{fig:btw_manna_sigma_a_r_2d}.
In the case of the BTW model the area and radius distributions 
display in contrast to the size and duration 
distribution the usual behavior.
The $q$-dependence of the exponents~$\sigma_a(q)$ 
and $\sigma_r(a)$ is given by a linear function 
and the slopes correspond to the trivial values
$\gamma_{ar}=2$ (compact avalanches) 
and $\gamma_{rr}=1$ (see Table~\ref{table:gamma}).
These exponents are the same for the Manna model
and the corresponding curves are parallel.
But as Fig.~\ref{fig:btw_manna_sigma_a_r_2d} shows
there is a clear shift between the curves of the
BTW and Manna model,
i.e., both models are characterized by two different
avalanche exponents $\tau_a$ and $\tau_r$.

The analysis of the perimeter distribution 
yields again a simple linear $q$-dependence
of the exponents~$\sigma_p(q)$ (not shown)
and we obtain $\gamma_{pr}=1.266\pm0.019$
which corresponds to the fractal dimension of the
boundary.
This value is in agreement 
with that obtained from a direct analysis of the 
scaling relation $p \sim r^{\gamma_{pr}}$ and 
differs significantly from the value of the
Manna model~$\gamma_{pr}=1.42$~\cite{BENHUR_1}.

Due to the nonlinear behavior of the size and 
duration distribution we use the area, radius,
and perimeter distributions in order to estimate
the intercept~$\Sigma$ of the BTW model
and obtain $\Sigma_{{\rm BTW},2D}=0.391 \pm 0.011$. 
But one has to be carefully to compare this 
result with the corresponding value of the 
Manna model.
The performed moment analysis based on the 
assumption that the scaling behavior
of the BTW model is given by a pure power-law
behavior [Eq.~(\ref{eq:power_law})].
But this assumption is in contradiction to
the observed logarithmic finite-size 
corrections~\cite{MANNA_1,LUEB_2} as well
as to recently reported investigations
where a multifractal behavior of the 
distribution was observed~\cite{DEMENECH_1}.
In the latter case no avalanche exponent 
could be defined whereas the analysis
of the logarithmic finite-size corrections
yield $\Sigma_{{\rm BTW},2D}\approx 2/3$~\cite{LUEB_2}
which again differs significantly from the 
corresponding value of the Manna model.

These results lead us to the conclusion that the 
moment analysis is applicable in order to
determine the geometric properties of the
avalanches of the BTW model,
e.g.~$\gamma_{ar}$, $\gamma_{pr}$.
The obtained results are in agreement with those
of previous investigations which based on different
methods of analyzing.
But in contrast to the geometric properties 
the moment analysis of the dynamical properties (size
and duration) of the avalanches exhibit
a non-trivial behavior.
This shows that the assumed simple power-law
behavior is not fulfilled and that scaling
corrections to Eq.~(\ref{eq:power_law})
cannot be neglected in these cases.
The question whether these corrections can be interpreted in terms of
unconventional finite-size effects~\cite{MANNA_1,LUEB_2},
or correspond to a cross-over effect from
the boundary to the bulk regime~\cite{ROBINSON_2,EDNEY_1},
or indicate a multifractal behavior
of the two-dimensional model~\cite{DEMENECH_1},
remains open.

\begin{figure}
 \epsfxsize=8.0cm
 \epsffile{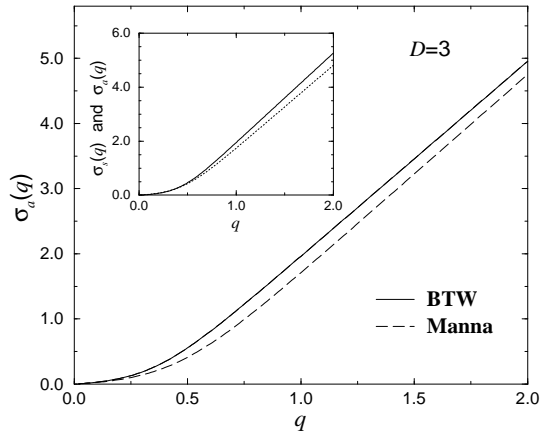}
 \caption{The exponent $\sigma_a(q)$  
          of the three-dimensional BTW and Manna model.
	  One obtains for both models the same asymptotic
	  slopes (see Table~\protect\ref{table:gamma}) 
	  which agree with $\gamma_{ar}=3$ (compact avalanches).
	  In the case of the BTW model both exponents
	  $\sigma_s(q)$ and $\sigma_a(q)$ are plotted, but
	  no difference between the curves can be seen.	 
	  The inset shows the exponents $\sigma_s(q)$ (solid line) 
	  and $\sigma_a(q)$ (dotted line)
	  for the Manna model. 
	  The different slopes indicate that multiple toppling
	  events are relevant in the case of the Manna model 
	  ($\gamma_{sa}=\gamma_{sr}/\gamma_{ar}>1$).
 \label{fig:btw_manna_sigma_a_3d}} 
\end{figure}

In the following we briefly compare the three-dimensional
BTW and Manna model and focus our attention
on the analysis of the size and area probability 
distribution.
In Fig.~\ref{fig:btw_manna_sigma_a_3d} we plot
the moment exponent $\sigma_a(q)$ for both models.
The result is similar to the two-dimensional 
case (Fig.~\ref{fig:btw_manna_sigma_a_r_2d}).
The slopes of the curves agree with the 
value~$\gamma_{ar}=3$ indicating that both models
are characterized by compact avalanches 
(see Table~\ref{table:gamma}).
But the two curves are shifted which shows that
the avalanche exponents~$\tau_a$ of the 
BTW ($\tau_a=1.352 \pm 0.022$) and
Manna ($\tau_a=1.436 \pm 0.018$) model are 
different for~$D=3$.

The determination of the intercepts $\Sigma$
confirms this result.
Analyzing for both models the size, area, duration
and radius distribution we obtain
$\Sigma_{{\rm BTW},3D}=1.016 \pm 0.056$
and $\Sigma_{{\rm Manna},3D}=1.333 \pm 0.036$.
The intercepts differs significantly.

\begin{table}[b]
\caption{Exponents $\gamma_{xr}$ for the BTW and Manna model. 
The values are obtained from a linear regression 
according to Eq.~(\protect\ref{eq:sigma_q_Sigma}) 
and the errors are of the order $\Delta \gamma_{xr}<0.01$.
The real error which includes the uncertainty of 
the whole moment analysis is hard to estimate and 
increases this value significantly (see text).}
\label{table:gamma}
\begin{tabular}{lcccc}
             	& BTW, $2D$	& Manna, $2D$	& BTW, $3D$	& Manna, $3D$\\  
\tableline 
$\gamma_{sr}$  	&        	& 2.764  	& 3.004    	& 3.302  \\ 
$\gamma_{ar}$  	& 2.021  	& 2.025  	& 3.004 	& 3.076  \\ 
$\gamma_{tr}$  	&        	& 1.540 	& 1.618         & 1.713  \\ 
$\gamma_{pr}$  	& 1.266  	& 1.42\protect\cite{BENHUR_1}	
						&               &        \\ 
$\gamma_{rr}$  	& 1.008		& 1.006		& 0.995         & 1.010  \\ 
\end{tabular}
\end{table}


Finally we consider the area and size distribution
for both models.
As mentioned above the three-dimensional Manna
model behaves different than the 
BTW model in the sense that multiple toppling
events affects the scaling behavior decisively.
This result can also be obtained from the moment
analysis. 
In the case of the BTW model no significant difference between 
the exponents $\sigma_a(q)$ and $\sigma_s(q)$ can be
observed~(see Fig.~\ref{fig:btw_manna_sigma_a_3d}).
Whereas for the Manna model both curves are characterized
by different slopes as can be seen in the inset of 
Fig.~\ref{fig:btw_manna_sigma_a_3d}.
A regression analysis yield $\gamma_{sr}=3.302 \pm 0.06$
and $\gamma_{ar}=3.076\pm 0.09$, respectively.
Thus multiple toppling events affect the scaling
behavior of the Manna model strongly ($\tau_s\neq \tau_a$),
in contrast to the BTW model.

\section{Summary}

In summary we performed a moment analysis of 
several probability distributions of the 
BTW and Manna model for $D=2$ and $D=3$.
We found that the corresponding moment 
exponents~$\sigma_x(q)$ differ significantly
for both models showing that the BTW and 
Manna model belong to different universality
classes.
Recently performed simulations of sandpile models
on a Sierpinski gasket display again the different
scaling behavior of both models~\cite{DAERDEN_1,DAERDEN_UN}.
Our results confirm the universality hypothesis
of Ben-Hur and Biham where the scaling behavior
of sandpiles models is determined by the way
in which the relaxing energy of critical sites
is distributed to the neighboring sites~\cite{BENHUR_1}.

I would like to thank D.\,V.~Ktitarev
for useful discussions
and a critical reading of 
the manuscript.


\begin{references} 
\bibitem[*]{SvenEmail} E-mail: sven@thp.uni-duisburg.de

\bibitem{BAK_1}
P. Bak, C. Tang, and K. Wiesenfeld, Phys. Rev. Lett. {\bf 59},  381  (1987).

\bibitem{BAK_2}
P. Bak, C. Tang, and K. Wiesenfeld, Phys.~Rev.~A {\bf 38},  364  (1988).

\bibitem{DHAR_2}
D. Dhar, Phys.~Rev.~Lett. {\bf 64},  1613  (1990).

\bibitem{MANNA_1}
{S.\,S.~Manna}, J.~Stat.~Phys. {\bf 59},  509  (1990).

\bibitem{GRASS_1}
P. Grassberger and S.~S. Manna, J.~Phys.~(France) {\bf 51},  1077  (1990).

\bibitem{CHRIS_2}
K. Christensen and Z. Olami, Phys.~Rev.~E {\bf 48},  3361  (1993).

\bibitem{ROBINSON_2}
{P.\,A.~Robinson}, Phys.~Rev.~E {\bf 49},  3919  (1994).

\bibitem{BENHUR_1}
A. Ben-Hur and O. Biham, Phys.~Rev.~E {\bf 53},  R1317  (1996).

\bibitem{LUEB_2}
S. L{\protect\"u}beck and {K.\,D.~Usadel}, Phys.~Rev.~E {\bf 55},  4095
  (1997).

\bibitem{EDNEY_1}
{S.\,D.~Edney}, {P.\,A.~Robinson}, and D. Chisholm, Phys.~Rev.~E {\bf 58},
  5395  (1998).

\bibitem{DEMENECH_1}
M. {De\,Menech}, {A.\,L.~Stella}, and C. Tebaldi, Phys.~Rev.~E {\bf 58},  2677
  (1998).

\bibitem{CHESSA_2}
A. Chessa, {H.\,E.~Stanley}, A. Vespignani, and S. Zapperi, Phys.~Rev.~E {\bf
  59},  12  (1999).

\bibitem{VAZQUEZ_1}
A. V{\'{a}}zquez and O. Sotolongo-Costa, cont-mat/9811414.

\bibitem{CHESSA_3}
A. Chessa, A. Vespignani, and S. Zapperi, cond-mat/9811365.

\bibitem{MANNA_2}
{S.\,S.~Manna}, J.~Phys.~A {\bf 24},  L363  (1991).

\bibitem{DHAR_7}
D. Dhar, Physica A {\bf 270}, 69  (1999).

\bibitem{DHAR_1}
D. Dhar and R. Ramaswamy, Phys. Rev. Lett. {\bf 63},  1659  (1989).

\bibitem{TADIC_3}
B. Tadi{\'{c}} {\it et~al.}, Phys.~Rev.~A {\bf 45},  8536  (1992).

\bibitem{LUEB_1}
S. L{\protect\"u}beck, B. Tadi{\'{c}}, and {K.\,D.~Usadel}, Phys.~Rev.~E {\bf
  53},  2182  (1996).

\bibitem{TADIC_1}
B. Tadi{\'{c}} and R. Ramaswamy, Physica~A {\bf 224},  188  (1996).

\bibitem{TADIC_2}
B. Tadi{\'{c}} and D. Dhar, Phys.~Rev.~Lett. {\bf 79},  1519  (1997).

\bibitem{PIETRO_3}
L. Pietronero, A. Vespignami, and S. Zapperi, Phys.~Rev.~Lett. {\bf 72},  1690
  (1994).

\bibitem{VESPIGNANI_3}
A. Vespignami, S. Zapperi, and L. Pietronero, Phys.~Rev.~E {\bf 51},  1711
  (1995).

\bibitem{KATORI_1}
M. Katori and H. Kobayashi, Physica {\bf 229},  461  (1996).

\bibitem{ZHANG_1}
Y.-C. Zhang, Phys. Rev. Lett. {\bf 63},  470  (1989).

\bibitem{DIAZ_1}
{A.~D\'{\i}az-Guilera}, Europhys.~Lett. {\bf 26},  177  (1994).

\bibitem{CORRAL_1}
{\'{A}.~Corral} and A. {D\'{\i}az-Guilera}, Phys.~Rev.~E {\bf 55},  2434
  (1997).

\bibitem{LUEB_3}
S. L{\protect\"u}beck, Phys.~Rev.~E {\bf 56},  1590  (1997).

\bibitem{DIAZ_3}
{A.~D\'{\i}az-Guilera}, Fractals {\bf 1},  963  (1993).

\bibitem{LUEB_4}
S. L{\protect\"u}beck and {K.\,D.~Usadel}, Phys.~Rev.~E {\bf 56},  5138
  (1997).

\bibitem{DAERDEN_1}
F. Daerden and C. Vanderzande, Physica A {\bf 256},  533  (1998).

\bibitem{DAERDEN_UN}
F. Daerden and C. Vanderzande (unpublished).


\end{references}
\end{document}